\documentclass{article}
\usepackage{frascatiphys}
\usepackage{graphicx}
\usepackage{psfrag}
\usepackage{graphics}
\usepackage{float}
\usepackage{amssymb,epsfig,amsmath,euscript,array}
\usepackage{cite}
\usepackage{multicol}
\usepackage{subfig}
\usepackage{fancyhdr}
\usepackage{color}
\usepackage{scalerel,stackengine}

\newcommand{\be}{\begin{equation}}
\newcommand{\ee}{\end{equation}}

\newcommand{\bm}[1]{\mbox{\boldmath $#1$}}

\def\bd{\begin{document}}
\def\ed{\end{document}}
\def\nn{\nonumber}
\def\bea{\begin{eqnarray}}
\def\eea{\end{eqnarray}}
\let\bm=\bibitem
\let\la=\label

\def\npb#1#2#3{Nucl. Phys. {\bf{B#1}} #3 (#2)}
\def\plb#1#2#3{Phys. Lett. {\bf{#1B}} #3 (#2)}
\def\prl#1#2#3{Phys. Rev. Lett. {\bf{#1}} #3 (#2)}
\def\prd#1#2#3{Phys. Rev. {D \bf{#1}} #3 (#2)}
\def\cmp#1#2#3{Comm. Math. Phys. {\bf{#1}} #3 (#2)}
\def\cqg#1#2#3{Class. Quantum Grav. {\bf{#1}} #3 (#2)}
\def\nppsa#1#2#3{Nucl. Phys. B (Proc. Suppl.) {\bf{#1A}}#3 (#2)}
\def\ap#1#2#3{Ann. of Phys. {\bf{#1}} #3 (#2)}
\def\ijmp#1#2#3{Int. J. Mod. Phys. {\bf{A#1}} #3 (#2)}
\def\rmp#1#2#3{Rev. Mod. Phys. {\bf{#1}} #3 (#2)}
\def\mpla#1#2#3{Mod. Phys. Lett. {\bf A#1} #3 (#2)}
\def\jhep#1#2#3{J. High Energy Phys. {\bf #1} #3 (#2)}
\def\atmp#1#2#3{Adv. Theor. Math. Phys. {\bf #1} #3 (#2)}

\newcommand{\EQ}[1]{\begin{equation} #1 \end{equation}}
\newcommand{\AL}[1]{\begin{subequations}\begin{align} #1 \end{align}\end{subequations}}
\newcommand{\SP}[1]{\begin{equation}\begin{split} #1 \end{split}\end{equation}}
\newcommand{\ALAT}[2]{\begin{subequations}\begin{alignat}{#1} #2 \end{alignat}\end{subequations}}
\def\beq{\begin{equation}}
\def\eeq{\end{equation}}

\def\N{{\cal N}}
\def\sst{\scriptscriptstyle}
\def\thetabar{\bar\theta}
\def\Tr{{\rm Tr}}
\def\one{\mbox{1 \kern-.59em {\rm l}}}
 \def\Nh{\hat{N}}

\def\a{\alpha}      \def\da{{\dot\alpha}}
\def\b{\beta}       \def\db{{\dot\beta}}
\def\c{\gamma}  \def\G{\Gamma}  \def\cdt{\dot\gamma}
\def\d{\delta}  \def\D{\Delta}  \def\ddt{\dot\delta}
\def\e{\epsilon}        \def\vare{\varepsilon}
\def\f{\phi}    \def\F{\Phi}    \def\vvf{\f}
\def\h{\eta}
\def\k{\kappa}
\def\l{\lambda} \def\L{\Lambda}
\def\m{\mu} \def\n{\nu}
\def\o{\omega}
\def\p{\pi} \def\P{\Pi}
\def\r{\rho}
\def\s{\sigma}  \def\S{\Sigma}
\def\t{\tau}
\def\th{\theta} \def\Th{\Theta} \def\vth{\vartheta}
\def\X{\Xeta}
\def\z{\zeta}

\def\cA{{\cal A}} \def\cB{{\cal B}} \def\cC{{\cal C}}
\def\cD{{\cal D}} \def\cE{{\cal E}} \def\cF{{\cal F}}
\def\cG{{\cal G}} \def\cH{{\cal H}} \def\cI{{\cal I}}
\def\cJ{{\cal J}} \def\cK{{\cal K}} \def\cL{{\cal L}}
\def\cM{{\cal M}} \def\cN{{\cal N}} \def\cO{{\cal O}}
\def\cP{{\cal P}} \def\cQ{{\cal Q}} \def\cR{{\cal R}}
\def\cS{{\cal S}} \def\cT{{\cal T}} \def\cU{{\cal U}}
\def\cV{{\cal V}} \def\cW{{\cal W}} \def\cX{{\cal X}}
\def\cY{{\cal Y}} \def\cZ{{\cal Z}}

\def\ua{\underline{\alpha}}
\def\ub{\underline{\phantom{\alpha}}\!\!\!\beta}
\def\uc{\underline{\phantom{\alpha}}\!\!\!\gamma}
\def\um{\underline{\mu}}
\def\ud{\underline\delta}
\def\ue{\underline\epsilon}
\def\una{\underline a}\def\unA{\underline A}
\def\unb{\underline b}\def\unB{\underline B}
\def\unc{\underline c}\def\unC{\underline C}
\def\und{\underline d}\def\unD{\underline D}
\def\une{\underline e}\def\unE{\underline E}
\def\unf{\underline{\phantom{e}}\!\!\!\! f}\def\unF{\underline F}
\def\unm{\underline m}\def\unM{\underline M}
\def\unn{\underline n}\def\unN{\underline N}
\def\unp{\underline{\phantom{a}}\!\!\! p}\def\unP{\underline P}
\def\unq{\underline{\phantom{a}}\!\!\! q}
\def\unQ{\underline{\phantom{A}}\!\!\!\! Q}
\def\unH{\underline{H}}

\def\As {{A \hspace{-6.4pt} \slash}\;}
\def\bs {{b \hspace{-6.4pt} \slash}\;}
\def\Ds {{D \hspace{-6.4pt} \slash}\;}
\def\ds {{\del \hspace{-6.4pt} \slash}\;}
\def\ss {{\s \hspace{-6.4pt} \slash}\;}
\def\ks {{ k \hspace{-6.4pt} \slash}\;}
\def\ps {{p \hspace{-6.4pt} \slash}\;}
\def\pas {{{p_1} \hspace{-6.4pt} \slash}\;}
\def\pbs {{{p_2} \hspace{-6.4pt} \slash}\;}

\def\Fh{\hat{F}}
\def\Vh{\hat{V}}
\def\Xh{\hat{X}}
\def\ah{\hat{a}}
\def\xh{\hat{x}}
\def\yh{\hat{y}}
\def\ph{\hat{p}}
\def\xih{\hat{\xi}}

\def\psit{\tilde{\psi}}
\def\Psit{\tilde{\Psi}}
\def\tht{\tilde{\th}}
\def\lt{\tilde{\lambda}}
\def\At{\tilde{A}}
\def\Qt{\tilde{Q}}
\def\Rt{\tilde{R}}
\def\Nt{\tilde{N}}

\def\at{\tilde{a}}
\def\st{\tilde{s}}
\def\ft{\tilde{f}}
\def\pt{\tilde{p}}
\def\qt{\tilde{q}}
\def\vt{\tilde{v}}
\def\nt{\tilde{n}}

\def\delb{\bar{\partial}}
\def\bz{\bar{z}}
\def\bD{\bar{D}}
\def\bB{\bar{B}}

\def\bk{{\bf k}}
\def\bl{{\bf l}}
\def\bp{{\bf p}}
\def\bq{{\bf q}}
\def\br{{\bf r}}
\def\bx{{\bf x}}
\def\by{{\bf y}}
\def\bR{{\bf R}}
\def\bV{{\bf V}}

\def\d{\delta}\def\D{\Delta}\def\ddt{\dot\delta}

\def\pa{\partial} \def\del{\partial}
\def\xx{\times}
\def\uno{\mbox{1 \kern-.59em {\rm l}}}

\def\trp{^{\top}}
\def\inv{^{-1}}
\def\dag{{^{\dagger}}}
\def\pr{^{\prime}}
\def\lan{\langle}
\def\ran{\rangle}

\def\rar{\rightarrow}
\def\lar{\leftarrow}
\def\lrar{\leftrightarrow}

\newcommand{\0}{\,\!}      %this is just NOTHING!
\def\one{1\!\!1\,\,}
\def\im{\imath}
\def\jm{\jmath}

\newcommand{\tr}{\mbox{tr}}
\newcommand{\slsh}[1]{/ \!\!\!\! #1}

\def\vac{|0\rangle}
\def\lvac{\langle 0|}

\def\hlf{\frac{1}{2}}
\def\ove#1{\frac{1}{#1}}

\def\Box{\square}
\def\ZZ{\mathbb{Z}}
\def\CC#1{({\bf #1})}
\def\bcomment#1{}
\def\bfhat#1{{\bf \hat{#1}}}
\def\VEV#1{\left\langle #1\right\rangle}

\newcommand{\ex}[1]{{\rm e}^{#1}} \def\ii{{\rm i}}

\def\rr{{\rm r}} \def\rs{{\rm s}}\def\rv{{\rm v}}
\def\ri{{\rm i}}\def\rj{{\rm j}}

\newcommand{\lrbrk}[1]{\left(#1\right)}
\newcommand{\sfrac}[2]{{\textstyle\frac{#1}{#2}}}

\newcommand\equalhat{\mathrel{\stackon[1.5pt]{=}{\stretchto{%
    \scalerel*[\widthof{=}]{\wedge}{\rule{1ex}{3ex}}}{0.5ex}}}}

\font\mybb=msbm10 at 12pt
\def\bb#1{\hbox{\mybb#1}}

\font\myBB=msbm10 at 18pt
\def\BB#1{\hbox{\myBB#1}}

\newcommand{\tclr}{\textcolor}
\newcommand{\bpmat}{\begin{pmatrix}}
\newcommand{\epmat}{\end{pmatrix}}
\newcommand{\mrm}[1]{\mathrm{#1}}
\newcommand{\mrs}[1]{\scriptscriptstyle{\mathrm{#1}}}
\newcommand{\vct}[1]{\boldsymbol{#1}}
\newcommand{\hf}{\frac{1}{2}}
\newcommand{\x}{\times}
\newcommand{\pd}{\partial}
\newcommand{\dslash}{\displaystyle{\not}}
\newcommand{\ol}[1]{\overline{#1}}
\newcommand{\abs}[1]{\vert{#1}\vert}
\newcommand{\chiSqM}{\chi^2_{\mrm{min}}}
\newcommand{\chiSqMDof}{\chi^2_{\mrm{min}}/\mrm{d.o.f.}}
\newcommand{\om}{\omega}
\newcommand{\Lag}{\mathcal{L}}
\newcommand{\ord}{\mathcal{O}}
\newcommand{\eps}{\epsilon}
\newcommand{\beFrac}{\frac{1-\be}{1+\be}}
\newcommand{\beFracI}{\frac{1+\be}{1-\be}}
\newcommand{\amu}{a_{\mu}}
\newcommand{\damu}{\delta\amu}
\newcommand{\Damu}{\Delta\amu}
\newcommand{\amuUnit}{10^{-10}}
\newcommand{\mmu}{m_{\mu}}
\newcommand{\amuQED}{\amu^{\mrm{QED}}}
\newcommand{\amuEW}{\amu^{\mrm{EW}}}
\newcommand{\amuEWl}{\amu^{\mrm{EW,}\,1l}}
\newcommand{\amuEWll}{\amu^{\mrm{EW,}\,2l}}
\newcommand{\amuh}{\amu^{\mrm{had}}}
\newcommand{\amuhLO}{\amu^{\text{had, LOVP}}}
\newcommand{\amuhHO}{\amu^{\text{had, HOVP}}}
\newcommand{\amuhHOa}{\amu^{\text{had, HOVP(a)}}}
\newcommand{\amuhHOb}{\amu^{\text{had, HOVP(b)}}}
\newcommand{\amuhHOc}{\amu^{\text{had, HOVP(c)}}}
\newcommand{\amuhLbL}{\amu^{\text{had, LbL}}}
\newcommand{\ff}[3]{\mathcal{F}_{\pi^{0{#1}}\gamma^{#2}\gamma^{#3}}}
\newcommand{\alps}{\alpha_s}
\newcommand{\asmz}{\alpha_s(M_Z^2)}
\newcommand{\amz}{\alpha(M_Z^2)}
\newcommand{\aqmz}{\alpha_{\mrm{QED}}(M_Z^2)}
\newcommand{\delAlp}{\Delta\alpha}
\newcommand{\dAlpL}{\delAlp_{\mrm{lep}}}
\newcommand{\dAlpT}{\delAlp_{\mrm{top}}}
\newcommand{\dAlpH}{\delAlp_{\mrm{had}}}
\newcommand{\dAlpHF}{\dAlpH^{(5)}}
\newcommand{\dAlpHFmz}{\dAlpHF(M_Z^2)}
\newcommand{\tmin}{t_{\mrm{min}}}
\newcommand{\sTh}{s_{\mrm{th}}}
\newcommand{\eTh}{\sqrt{\sTh}}
\newcommand{\Ekmi}{E^{\,(k,m)}_i}
\newcommand{\Nkm}{N^{(k,m)}}
\newcommand{\Nkn}{N^{(k,n)}}
\newcommand{\Nexp}{N_{\mrm{exp}}}
\newcommand{\Nclu}{N_{\mrm{clu}}}
\newcommand{\Ntot}{N_{\mrm{tot}}}
\newcommand{\Rkmi}{R^{\,(k,m)}_i}
\newcommand{\Rknj}{R^{\,(k,n)}_j}
\newcommand{\dRkmi}{\mrm{d}\Rkmi}
\newcommand{\dRtkmi}{\mrm{d}\tilde{R}^{\,(k,m)}_i}
\newcommand{\BR}[2]{\mathcal{B}(#1\to #2)}
\newcommand{\decay}[2]{#1\to #2}
\newcommand{\UpsIVs}{\Upsilon(4S)}
\newcommand{\Gee}{\Gamma_{ee}}
\newcommand{\Gtot}{\Gamma_{\mrm{tot}}}
\newcommand{\ppC}{\pi^+\pi^-}
\newcommand{\ppN}{\pi^0\pi^0}
\newcommand{\pppC}{\pi^+\pi^-\pi^0}
\newcommand{\kkC}{K^+K^-}
\newcommand{\kskl}{K^0_S K^0_L}
\newcommand{\ksks}{K^0_S K^0_S}
\newcommand{\klkl}{K^0_L K^0_L}
\newcommand{\kskp}{K^0_S K^{\pm}\pi^{\mp}}
\newcommand{\eeMuMu}{e^+e^-\to\mu^+\mu^-}
\newcommand{\eeHadr}{e^+e^-\to\mrm{hadrons}}
\newcommand{\eeGhadr}{e^+e^-\to\gamma^*\to\mrm{hadrons}}
\newcommand{\tauNuHadr}{\tau\to\nu_{\tau}+\mrm{hadrons}}
\newcommand{\eeGPiPi}{e^+e^-\to\gamma^*\to\pi^+\pi^-}
\newcommand{\tauNuWNuPiPi}{\tau\to\nu_{\tau}W\to\nu_{\tau}\pi\pi^0}
\newcommand{\eeGIncl}{e^+e^-\to\gamma^*\to\mrm{all\,hadrons}}
\newcommand{\eeIncl}{e^+e^-\to\mrm{all\,hadrons}}
\newcommand{\eePiG}{e^+e^-\to\pi^0\gamma}
\newcommand{\eePiPi}{e^+e^-\to\pi^+\pi^-}
\newcommand{\eePiPiPi}{e^+e^-\to\pi^+\pi^-\pi^0}
\newcommand{\eeKK}{e^+e^-\to K^+K^-}
\newcommand{\ch}{\mrm{ch}}
\newcommand{\iso}{\mrm{iso}}
\newcommand{\noeta}{\text{no }\eta}
\newcommand{\kkr}{K\bar{K}\rho}
\newcommand{\kkp}{K\bar{K}\pi}
\newcommand{\kkpp}{K\bar{K}2\pi}
\newcommand{\kkppp}{K\bar{K}3\pi}
\newcommand{\isoAA}{(2\pi^+2\pi^-\pi^0)_{\mrm{no}\,\eta}}
\newcommand{\isoAB}{(\pi^+\pi^-3\pi^0)_{\mrm{no}\,\eta}}
\newcommand{\isoAC}{\omega(\to\mrm{npp})2\pi}
\newcommand{\isoACf}{\omega(\to\text{non-pure pionic states})2\pi}
\newcommand{\isoAD}{\eta\pi^+\pi^-}
\newcommand{\isoBA}{(2\pi^+2\pi^-2\pi^0)_{\mrm{no}\,\eta}}
\newcommand{\isoBB}{(\pi^+\pi^-4\pi^0)_{\mrm{no}\,\eta}}
\newcommand{\isoBC}{3\pi^+3\pi^-}
\newcommand{\isoBD}{\omega(\to\mrm{npp})3\pi}
\newcommand{\isoBDf}{\omega(\to\text{non-pure pionic state})3\pi}
\newcommand{\isoBE}{\eta\omega}
\newcommand{\isoEA}{\kkppp}
\newcommand{\isoEAa}{(K^+K^-\pi^+\pi^-\pi^0)_{\mrm{no}\,\eta}}
\newcommand{\isoEAb}{(K^0\bar{K}^0\pi^+\pi^-\pi^0)_{\mrm{no}\,\eta}}
\newcommand{\isoEB}{\omega(\to\mrm{npp})K\bar{K}}
\newcommand{\isoEBf}{\omega(\to\text{non-pure pionic states})K\bar{K}}
\newcommand{\isoEC}{\eta\phi}
\newcommand{\isoFA}{\eta2\pi^+2\pi^-}
\newcommand{\isoFB}{\eta\pi^+\pi^-2\pi^0}
\newcommand{\sigEEhadr}{\sigma(\eeHadr)}
\newcommand{\sigHad}{\sigma_{\mrm{had}}}
\newcommand{\sigHadB}{\sigHad^0}
\newcommand{\sigPt}{\sigma_{\mrm{pt}}}
\newcommand{\Rhad}{R_{\mrm{had}}}
\begin{document}
\title{ 
THE HADRONIC VACUUM POLARISATION CONTRIBUTIONS TO THE MUON {\large$g-2$}
}
\author{
Alexander Keshavarzi       \\
{\em Department of Mathematical Sciences, University of Liverpool, Liverpool L69 3BX, UK} \\
Daisuke Nomura        \\
{\em KEK Theory Center, Tsukuba, Ibaraki 305-0801, Japan} \\
{\em Yukawa Institute for Theoretical Physics, Kyoto University, Kyoto 606-8502, Japan} \\
Thomas Teubner      \\
{\em Department of Mathematical Sciences, University of Liverpool, Liverpool L69 3BX, UK} \\
}
\maketitle
\baselineskip=10pt
\begin{abstract}
The hadronic vacuum polarisation contributions to the anomalous magnetic moment of the muon, $a_{\mu}^{\rm had, VP}$ have been re-evaluated from the combination of $e^+e^-\rightarrow {\rm hadrons}$ cross section data. Focus has been placed on the development of a new data combination method, which fully incorporates all correlated statistical and systematic uncertainties in a bias free approach. Using these combined data have resulted in estimates of the hadronic vacuum polarisation contributions to $g-2$ of the muon of $a_{\mu}^{\rm had, \, LO \, VP} = (693.27 \pm 2.46)\times 10^{-10}$ and $a_{\mu}^{\rm had, \, NLO \, VP} = (-9.82 \pm 0.04)\times 10^{-10}$. The new estimate for the Standard Model prediction is found to be $a_{\mu}^{\rm SM}  =  (11\ 659 \ 182.05  \pm 3.56) \times 10^{-10}$, which is $3.7\sigma$ below the current experimental measurement. In addition, the prediction for the hadronic contribution to the QED coupling at the $Z$ boson mass has been calculated to be  $\Delta\alpha_{\rm had}^{(5)}(M_Z^2)= (276.11 \pm 1.11)\times 10^{-4}$, resulting in $\alpha^{-1}(M_Z^2) = 128.946 \pm 0.015$. 
\end{abstract}
\baselineskip=14pt

\section{Introduction}
The anomalous magnetic moment of the muon, $a_{\mu} = (g-2)_{\mu}/2$,
stands as an enduring test of the Standard Model (SM), where the
$\sim3.5\sigma$  (or higher) discrepancy between the experimental
measurement $a_{\mu}^{\rm exp} = 11\ 659 \ 209.1 \ (5.4) \ (3.3)  \times 10^{-10}$~\cite{Bennett:2002jb,PDG2016}  and the SM prediction 
$a_{\mu}^{\rm SM}$ could be an indication of the existence of new
physics beyond the SM. Efforts to improve the experimental estimate at Fermilab
(FNAL)~\cite{Grange:2015fou} and at J-PARC~\cite{Mibe:2010zz} aim to
reduce the experimental uncertainty by a factor of four compared to
the BNL measurement. It is therefore imperative that the SM prediction
is also improved to determine whether the $g-2$ discrepancy is well
established. 

The uncertainty of $a_{\mu}^{\rm SM}$ is completely dominated by the
hadronic contributions, where the hadronic vacuum polarisation contributions can be
separated into the leading-order (LO) and higher-order contributions. These are calculated utilising dispersion integrals
and the experimentally measured cross section $\sigma^0_{{\rm had},\gamma} (s) \equiv \sigma^0(e^+e^-\rightarrow
\gamma^* \rightarrow {\rm hadrons} + \gamma)$, where the superscript 0 denotes the bare cross section (undressed of
all vacuum polarisation (VP) effects) and the subscript $\gamma$ indicates
the inclusion of effects from final state photon radiation (FSR). At
LO, the dispersion relation reads 
\beq \label{eq:amu}
a_{\mu}^{\rm had,\,LO\,VP} =
\frac{\alpha^2}{3\pi^2}\int^{\infty}_{m_{\pi}^2} \frac{{\rm d}s}{s}
R(s)K(s) \ \ \ \ ; \ \ \ \ R(s) = \frac{\sigma^0_{{\rm had},\gamma} (s)}{\sigma_{\rm pt}(s)}
\equiv \frac{\sigma^0_{{\rm had},\gamma} (s)}{4\pi\alpha^2/3s} \, ,
\eeq
where $K(s)$ is a well known kernel function. In addition to calculating $a_{\mu}^{\rm had,\,VP}$, the combination of
hadronic cross section data is also used to calculate the effective QED coupling
at the $Z$ boson mass, $\alpha(M_{Z}^2)$, which is the least precisely
known of the three fundamental electro-weak (EW) parameters of the SM
(the Fermi constant $G_{\mu}$, $M_Z$ and $\alpha(M_{Z}^2)$) and
hinders the accuracy of EW precision fits.

\section{Radiative corrections and data combination}

Equation \eqref{eq:amu} requires the
experimental cross section to be undressed of all
VP effects. However, recent data are more commonly undressed 
in the experimental analyses already, removing the need
to apply a correction to these data sets and, hence, reducing the impact of
the extra radiative correction uncertainty which is applied to each channel. Concerning FSR, detailed studies
have been performed for the important $\pi^+\pi^-$ and $K^+ K^-$ channels.
The $K^+K^-$ final state is dominated by the $\phi$ peak, where the phase space for real radiation is
severely restricted and the possibility for any hard real radiation is strongly suppressed. Therefore,
no correction or additional error estimate due to FSR is now applied
in the $K^+K^-$ channel (or the $K^0_S K^0 _L$ channel).  For the two pion channel, in principle larger contributions
from real radiation can arise. Therefore, the fully inclusive scalar QED correction~\cite{Hoefer:2001mx} is applied  where necessary. It should be noted, however, that recent sets from radiative
return (where accounting for FSR effects is an integral part of the analysis) have now become dominant in the $\pi^+ \pi^-$ data combination, 
reducing the impact of the fully inclusive FSR
correction from older data. 
For the sub-leading, multi-hadron channels, there are, at present, no
equivalent FSR calculations. Therefore, possible
effects are accounted for by applying a conservative additional uncertainty.

Within each hadronic channel, data points from different experiments
are re-binned into {\em clusters}~\cite{Keshavarzi:2018mgv}. A covariance matrix is then constructed for the combination which contains
the uncertainty and correlation information of all data
points. Using the covariance
matrix as defined by the data alone could result in bias
(see~\cite{D'Agostini:1993uj,Blobel:2003wa}). To avoid this, the covariance matrix is redefined at each step of an iterative linear $\chi^2$-minimisation~\cite{Ball:2009qv,Benayoun:2015gxa} using the fitted
values for the cluster centres, $R_m$ (see~\cite{Keshavarzi:2018mgv}). Convergence of the iteration is observed in this work to
occur after only a few steps. Performing the minimisation yields the cluster centres $R_m$ and the
covariance matrix $V\big(m,n\big)$, which is inflated according to the local
$\chi^2_{\rm min}/{\rm d.o.f.}$ for each cluster if $\chi^2_{\rm min}/{\rm
  d.o.f.} > 1$. This is done in order to account for any tensions
between the data. The use of the full covariance matrix allows for the inclusion of
any-and-all uncertainties and correlations that may exist between the
measurements. Hence, the appropriate influence of the
correlations is incorporated into the determination of the cluster
centres, $R_m$, with the correct propagation of all experimental
errors to the uncertainty. The data
are then integrated in order to obtain
$a_{\mu}^{\rm had, VP}$ and the five flavour contribution to the running $\alpha$, $\Delta \alpha_{\rm had}^{(5)}(M_Z^2)$.

\section{Determining $a_{\mu}^{\rm had, VP}$ and $\Delta\alpha_{\rm had}^{(5)}(M_Z^2)$}\label{amuanddeltaalpha}

\begin{figure}[!t] 
  \centering
    {\includegraphics[width=0.6\textwidth]{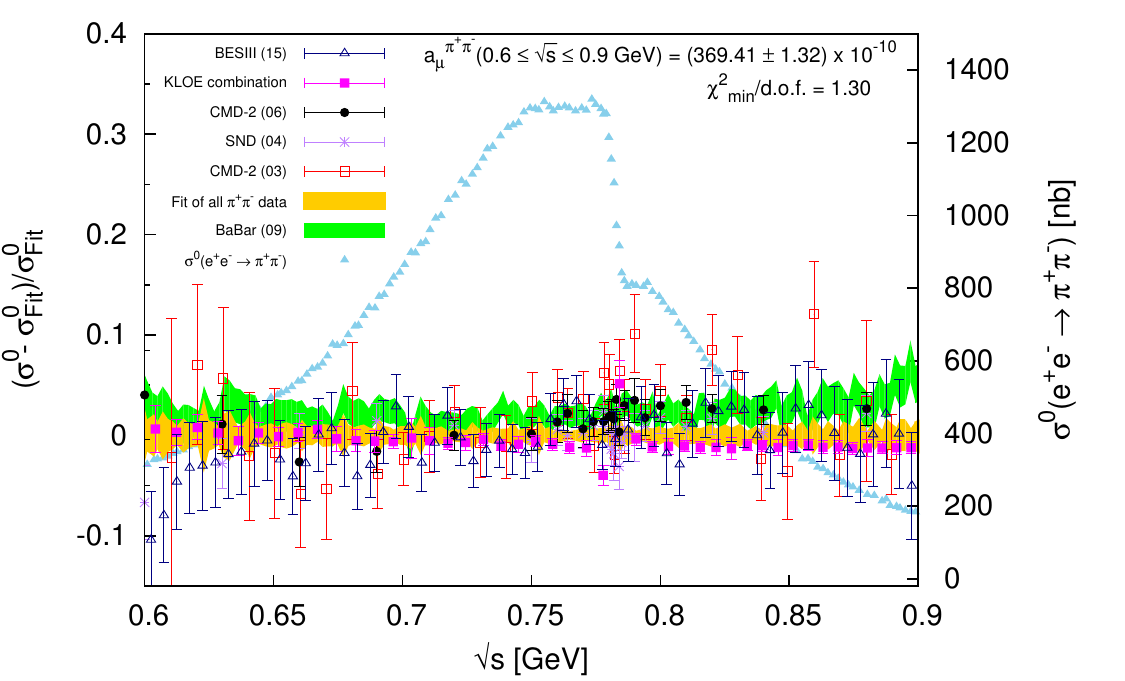}
     \caption{The relative difference of the radiative return and most relevant direct scan data sets contributing to $a_{\mu}^{\pi^+\pi^-}$ and the combination of all data, plotted in the $\rho$ region. The width of the coloured bands represent the propagation of the statistical and systematic uncertainties added in quadrature.}     \label{fig:RadRetFit}}
\end{figure}
\begin{figure}[!t]
\centering
  \subfloat[$\sigma^{0}(e^+e^-\rightarrow\pi^+\pi^-)$]{%
    \includegraphics[width= 0.33\textwidth]{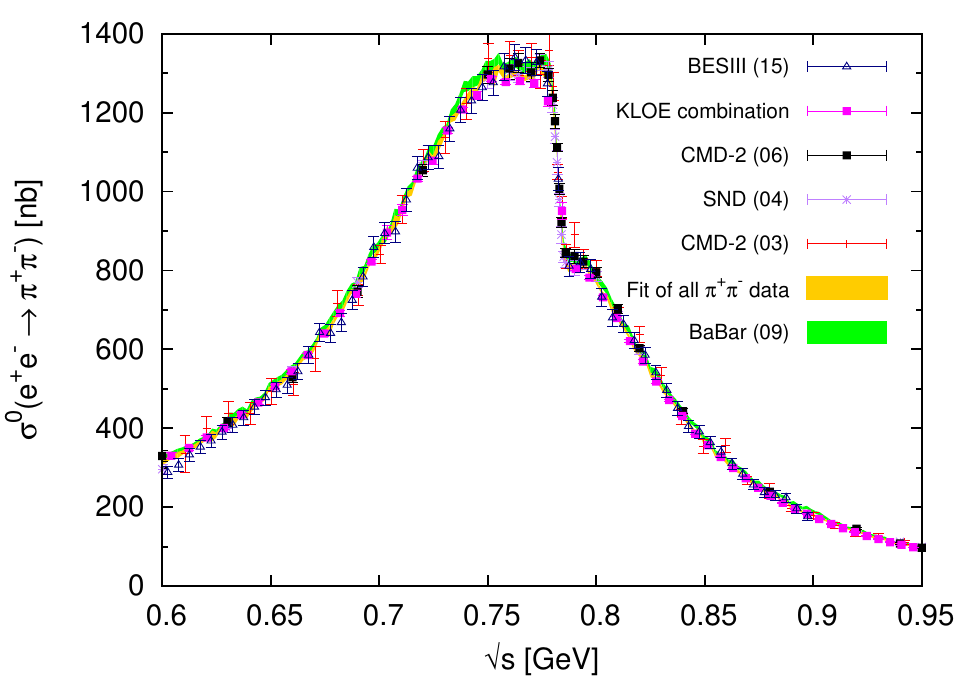}}\hfill
  \subfloat[$\sigma^{0}(e^+e^-\rightarrow\pi^+\pi^-\pi^0)$]{%
    \includegraphics[width= 0.33\textwidth]{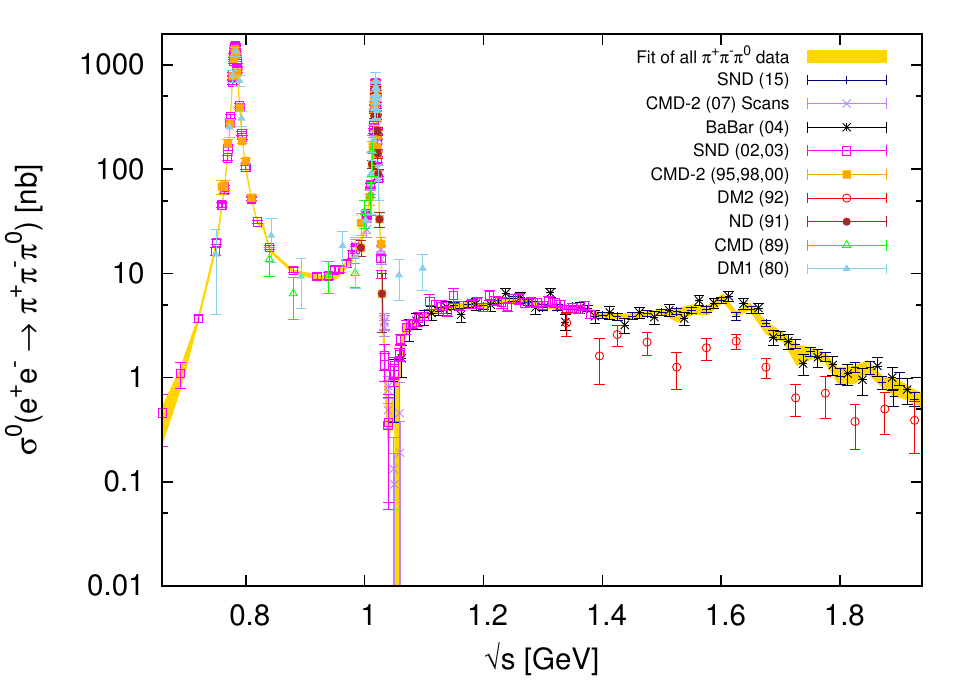}}\hfill
  \subfloat[$\sigma^{0}(e^+e^-\rightarrow\pi^+\pi^-\pi^+\pi^-)$]{%
    \includegraphics[width= 0.33\textwidth]{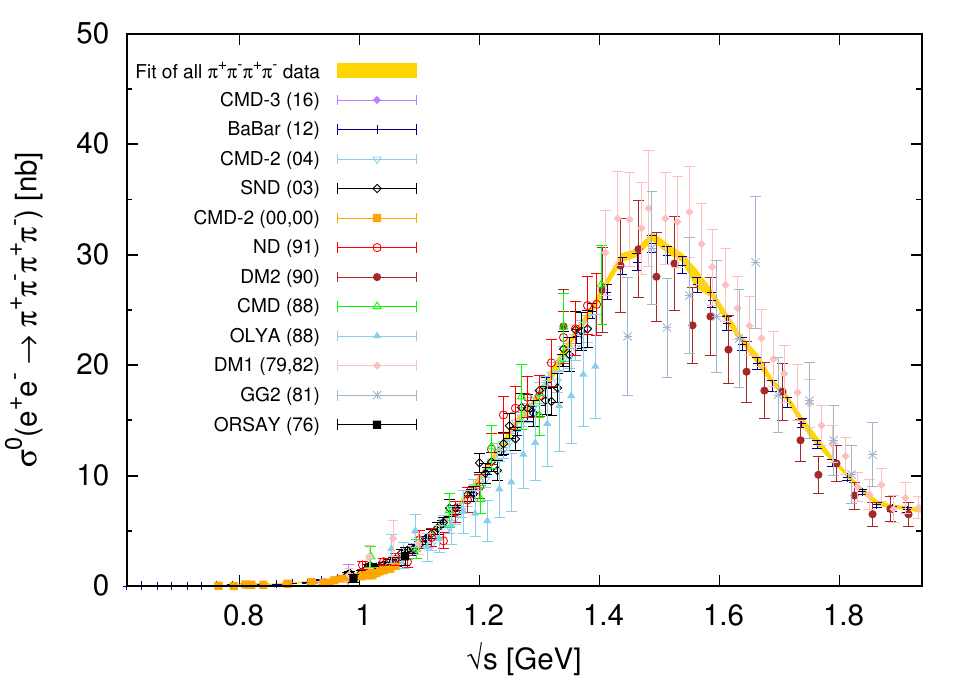}}\hfill
  \subfloat[$\sigma^{0}(e^+e^-\rightarrow\pi^+\pi^-\pi^0\pi^0)$]{%
    \includegraphics[width= 0.33\textwidth]{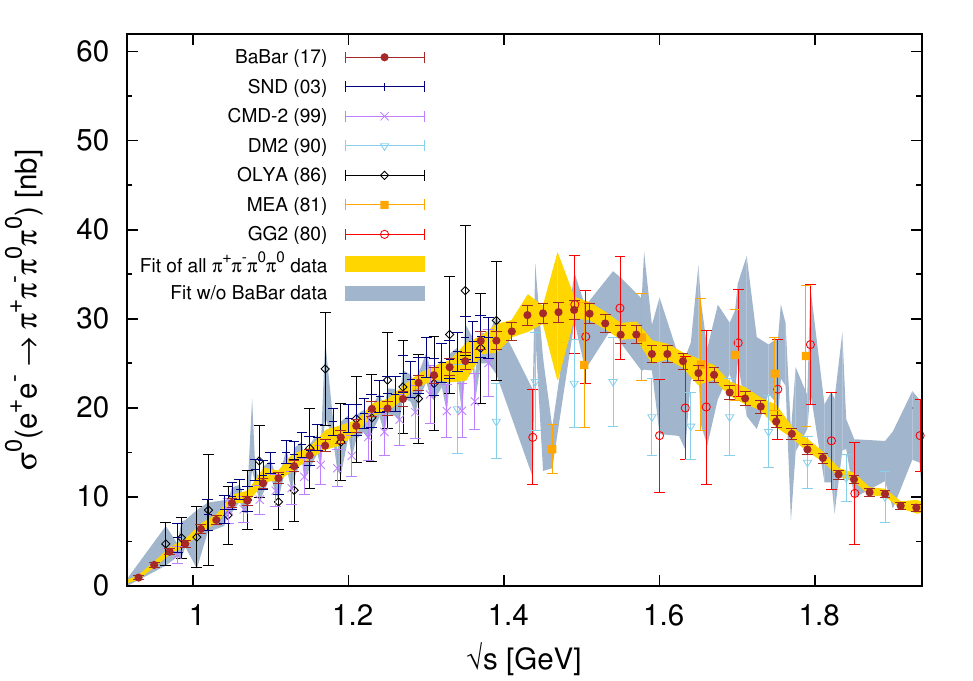}}\hfill
  \subfloat[$\sigma^{0}(e^+e^-\rightarrow K^+K^-)$]{%
    \includegraphics[width= 0.33\textwidth]{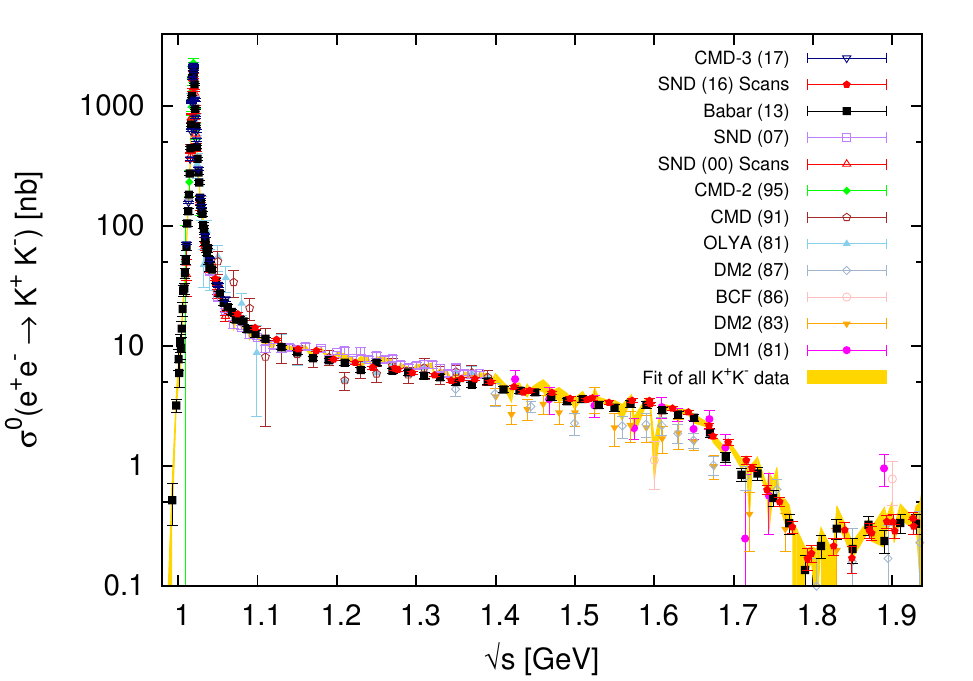}}\hfill
  \subfloat[$\sigma^{0}(e^+e^-\rightarrow K^0_S K^0_L)$]{%
    \includegraphics[width= 0.33\textwidth]{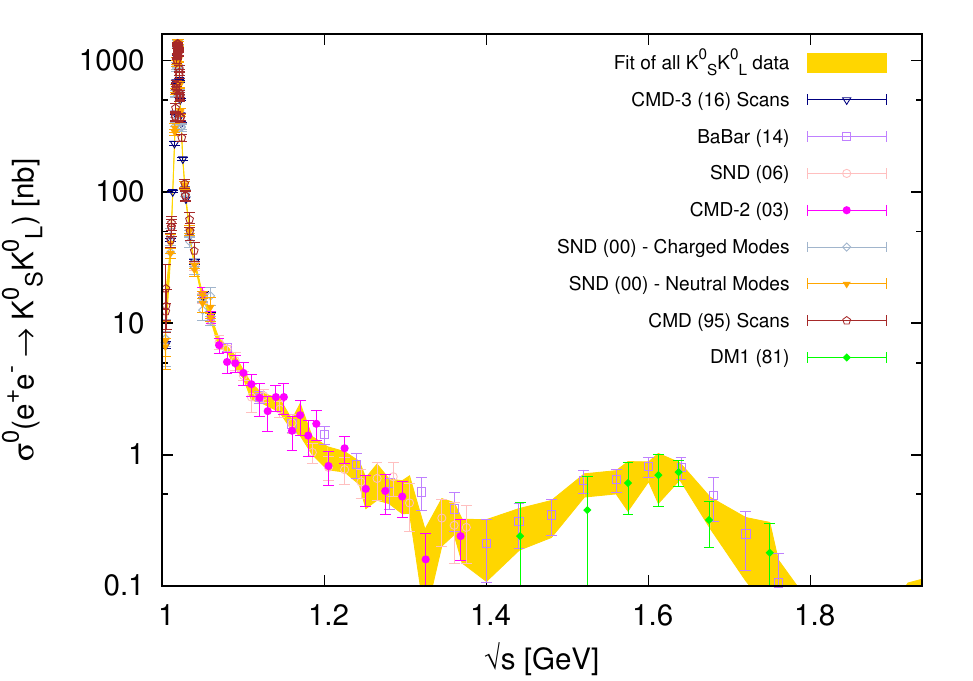}}\hfill
  \subfloat[$\sigma^{0}(e^+e^-\rightarrow KK\pi)$]{%
    \includegraphics[width= 0.33\textwidth]{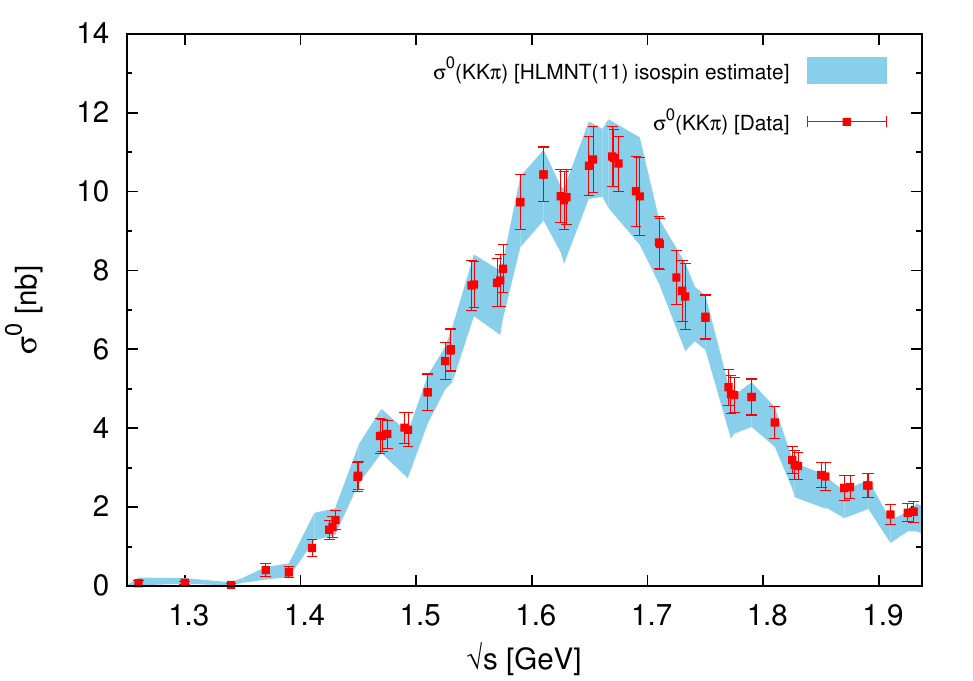}}\hfill
  \subfloat[$\sigma^{0}(e^+e^-\rightarrow KK\pi\pi)$]{%
    \includegraphics[width= 0.33\textwidth]{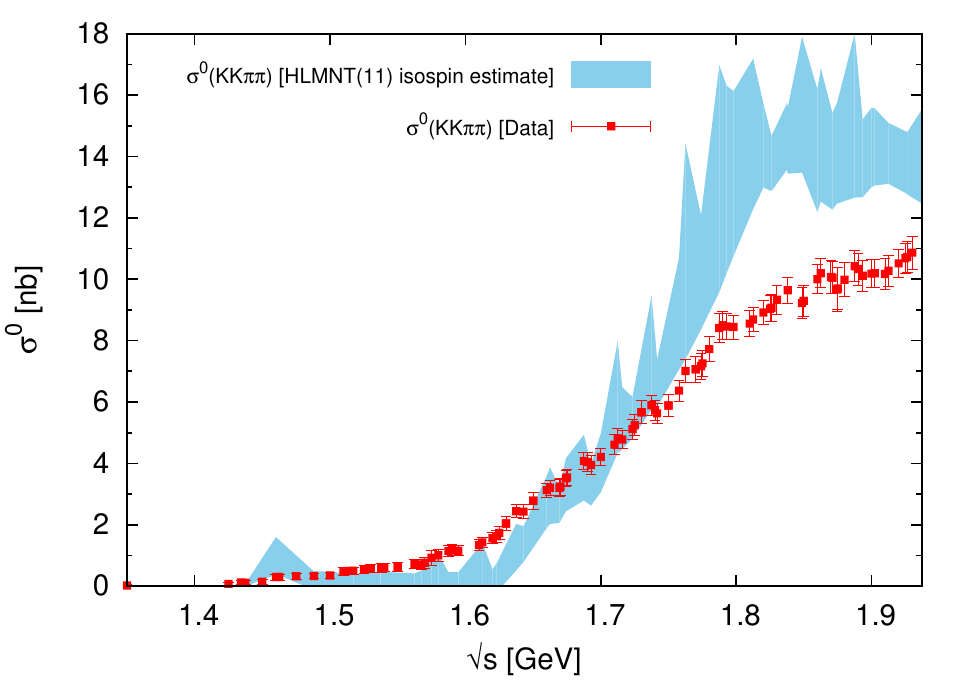}}\hfill
  \subfloat[Inclusive data]{%
    \includegraphics[width= 0.33\textwidth]{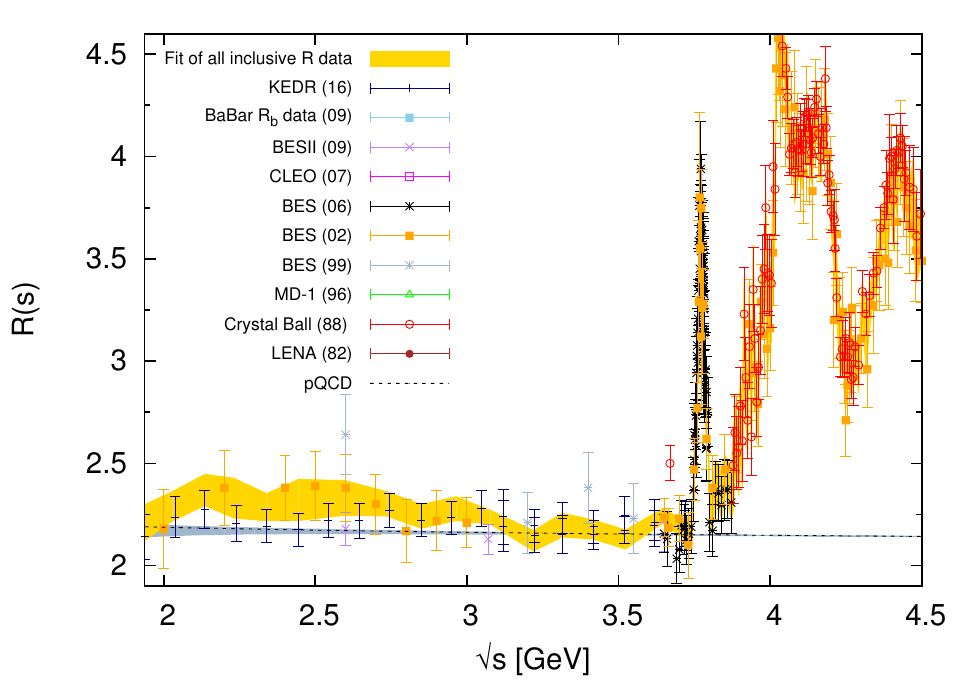}}\hfill
  \caption{The resulting cross sections of the leading and major sub-leading hadronic final states.}\label{fig:excxSec}
\end{figure}

 
In the  $\pi^+\pi^-$ channel, two recent radiative return measurements from the KLOE collaboration~\cite{Babusci:2012rp,KLOEcombination} and the BESIII collaboration~\cite{Ablikim:2015orh} in the $\rho$ region have improved the estimate of this final state. As indicated in Figure~\ref{fig:RadRetFit}, tension exists between the BaBar data~\cite{Aubert:2009ad} and all other contributing data. Although BaBar still influences the combination with an increase, the agreement between the other radiative return measurements and the direct scan data largely compensates for this effect. However, the tension between data is reflected in the local $\chi^2_{\rm min}/{\rm d.o.f.}$ error inflation, which results in an $\sim15\%$ increase in the uncertainty of $a_{\mu}^{\pi^+\pi^-}$.
The full combination of all $\pi^+\pi^-$ data results in the contributions to $a_{\mu}^{\rm had, \, LO \, VP}$ and $\Delta\alpha_{\rm had}^{(5)}(M_Z^2)$ as given in Table~\ref{tab:amuhadexc}. The cross section in the $\rho$ region is displayed in plot (a) of Figure~\ref{fig:excxSec}.
\begin{table}[!t]
\centering
%\vspace{-1cm}
%\hspace{2.0cm}
\scalebox{0.9}{
 {\renewcommand{\arraystretch}{0.9}
 \begin{tabular}{|l|c|c|c|}
\hline																						
{\bf Channel}	&	{\bf Energy range (GeV)}							&	$a_{\mu}^{\rm had, \, LO \, VP} \times 10^{10}$					&	$\Delta\alpha_{\rm had}^{(5)}(M_Z^2) \times 10^{4}$	\\
\hline			
$\pi^+\pi^-$	&	$	0.305	\leq	\sqrt{s}	\leq	1.937	$	&	$	502.97	\pm	1.97\hphantom{0}	$	&	$	34.26	\pm	0.12	$		\\								$\pi^+\pi^-\pi^0$	&	$	0.660	\leq	\sqrt{s}	\leq	1.937	$	&	$	47.79	\pm	0.89	$	&	$	\hphantom{0}4.77	\pm	0.08	$		\\
$\pi^+\pi^-\pi^+\pi^-$	&	$	0.613	\leq	\sqrt{s}	\leq	1.937	$	&	$	14.87	\pm	0.20	$	&	$	\hphantom{0}4.02	\pm	0.05	$	\\
$\pi^+\pi^-\pi^0\pi^0$	&	$	0.850	\leq	\sqrt{s}	\leq	1.937	$	&	$	19.39	\pm	0.78	$	&	$	\hphantom{0}5.00	\pm	0.20	$		\\
$K^+K^-$	&	$	0.988	\leq	\sqrt{s}	\leq	1.937	$	&	$	23.03	\pm	0.22	$	&	$	\hphantom{0}3.37	\pm	0.03	$		\\
$K^0_S K^0_L$	&	$	1.004	\leq	\sqrt{s}	\leq	1.937	$	&	$	13.04	\pm	0.19	$	&	$	\hphantom{0}1.77	\pm	0.03	$		\\
$KK\pi$	&	$	1.260	\leq	\sqrt{s}	\leq	1.937	$	&	$	\hphantom{0}2.71	\pm	0.12	$	&	$	\hphantom{0}0.89	\pm	0.04	$	\\
$KK2\pi$	&	$	1.350	\leq	\sqrt{s}	\leq	1.937	$	&	$	\hphantom{0}1.93	\pm	0.08	$	&	$	\hphantom{0}0.75	\pm	0.03	$		\\
Inclusive channel	&	$	1.937	\leq	\sqrt{s}	\leq	11.200	$	&	$	43.67	\pm	0.67	$	&	$	82.82	\pm	1.05	$		\\
\hline																						\end{tabular} 
} 
}\caption{Contributions to $a_{\mu}^{\rm had, \, LO \, VP}$ and $\Delta\alpha_{\rm had}^{(5)}(M_Z^2)$~\cite{Keshavarzi:2018mgv}. The first column indicates the hadronic final state or individual contribution, the second column gives the respective energy range of the contribution, the third column gives the determined value of $a_{\mu}^{\rm had, \, LO \, VP}$ and the last column states the value of $\Delta\alpha_{\rm had}^{(5)}(M_Z^2)$.}\label{tab:amuhadexc}
\end{table}

The results and cross sections from other major sub-leading channels are also given in Table~\ref{tab:amuhadexc} and Figure~\ref{fig:excxSec}, respectively. In all cases, these channels include new data sets which, coupled with the new data combination routine, have improved the estimates of $a_{\mu}^{\rm had, \, LO \, VP}$ and $\Delta \alpha_{\rm had}^{(5)}(M_Z^2)$ from these final states. In particular, a new measurement of the $\pi^+\pi^-\pi^0\pi^0$ channel by BaBar~\cite{TheBaBar:2017vzo} has provided the only new data in this channel since 2003. The uncertainty contribution from $\pi^+\pi^-\pi^0\pi^0$ is, however, still relatively large in comparison with its contribution to $a_{\mu}^{\rm had, \, LO \, VP}$ and requires better new data. Notably, the $K^+K^-$ channel now includes a precise and finely binned measurement by the BaBar collaboration,  supplemented with full statistical and systematic covariance matrices~\cite{Lees:2013gzt}, being the first and only example to date of the release of energy dependent, correlated uncertainties outside of the $\pi^+\pi^-$ channel. The neutral final state $K^0_S K^0_L\pi^0$ has been measured by SND~\cite{Achasov:2017vaq} and BaBar~\cite{TheBABAR:2017vgl}, completing all modes that contribute to the $KK\pi$ final state. Plot (g) of Figure~\ref{fig:excxSec} demonstrates good agreement between the previously used isospin estimate~\cite{Hagiwara:2011af} and the data-based approach in this analysis. In addition, BaBar have also completed all modes that contribute to the $KK\pi\pi$ channel~\cite{TheBABAR:2017vgl}. Examining plot (h) of Figure~\ref{fig:excxSec}, it is evident that the isospin relations provided a poor estimate of this final state.
The inclusive hadronic $R$-ratio now includes precise measurements by the KEDR collaboration~\cite{Anashin:2016hmv}. The fit of the inclusive data in the range $1.937\leq\sqrt{s}\leq 3.80$ GeV is shown in plot (i) of Figure~\ref{fig:excxSec}, which demonstrates that the inclusive data combination is much improved. With the new KEDR data, the differences between the inclusive data and pQCD are not as large as previously and, hence, the contributions in the entire inclusive data region are now estimated using the inclusive data alone. 

\subsection{Total contribution of $a_{\mu}^{\rm had, \, LO \, VP}$ and $\Delta\alpha_{\rm had}^{(5)}$}
 \begin{figure}[!t]
\centering
  \subfloat[Fractional contributions to $a_{\mu}^{\rm had, \, LO \, VP}$]{%
    \includegraphics[width= 0.5\textwidth]{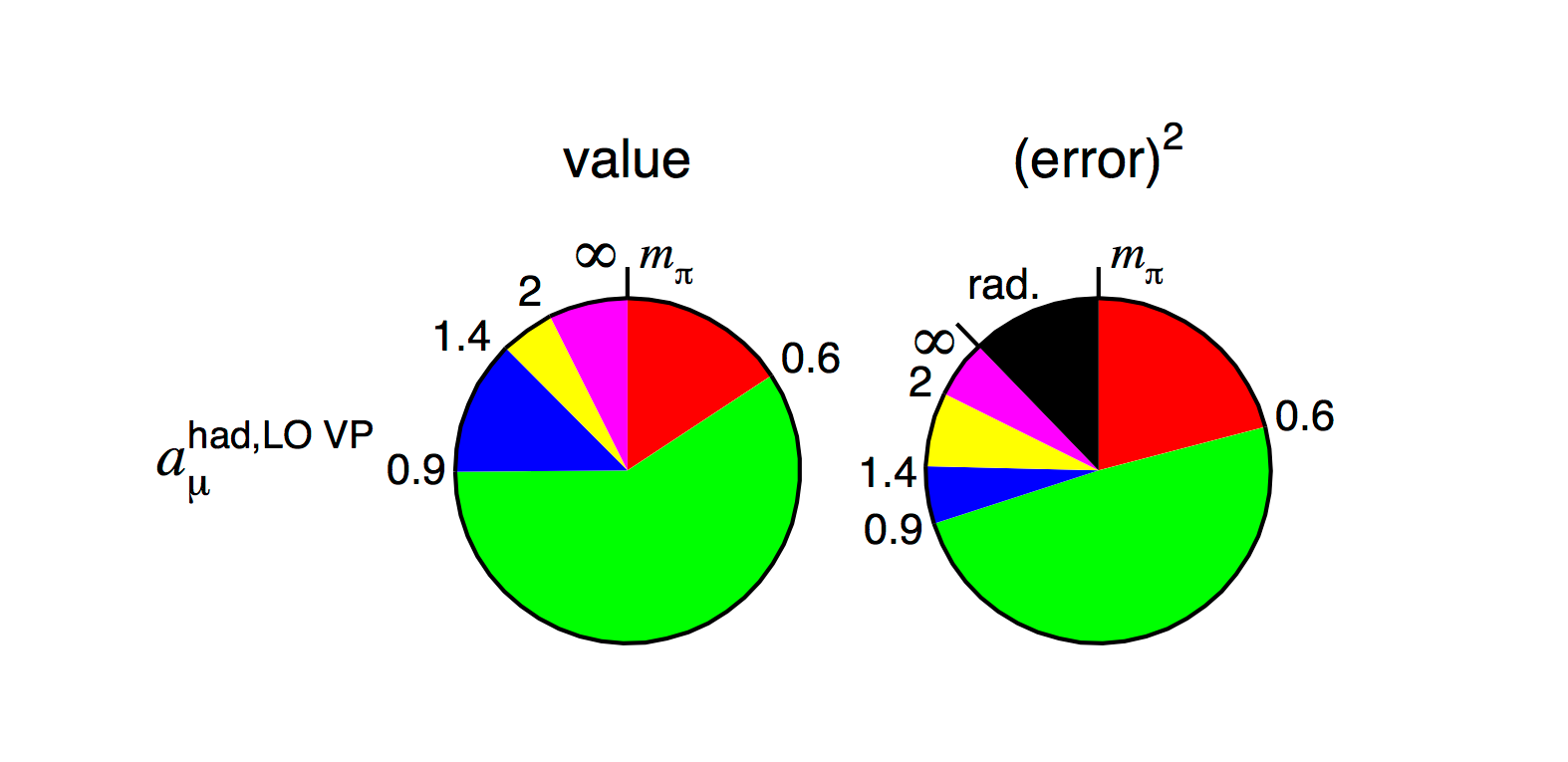}}\hfill
  \subfloat[Fractional contributions to $\Delta\alpha_{\rm had}^{(5)}(M_Z^2)$]{%
    \includegraphics[width= 0.5\textwidth]{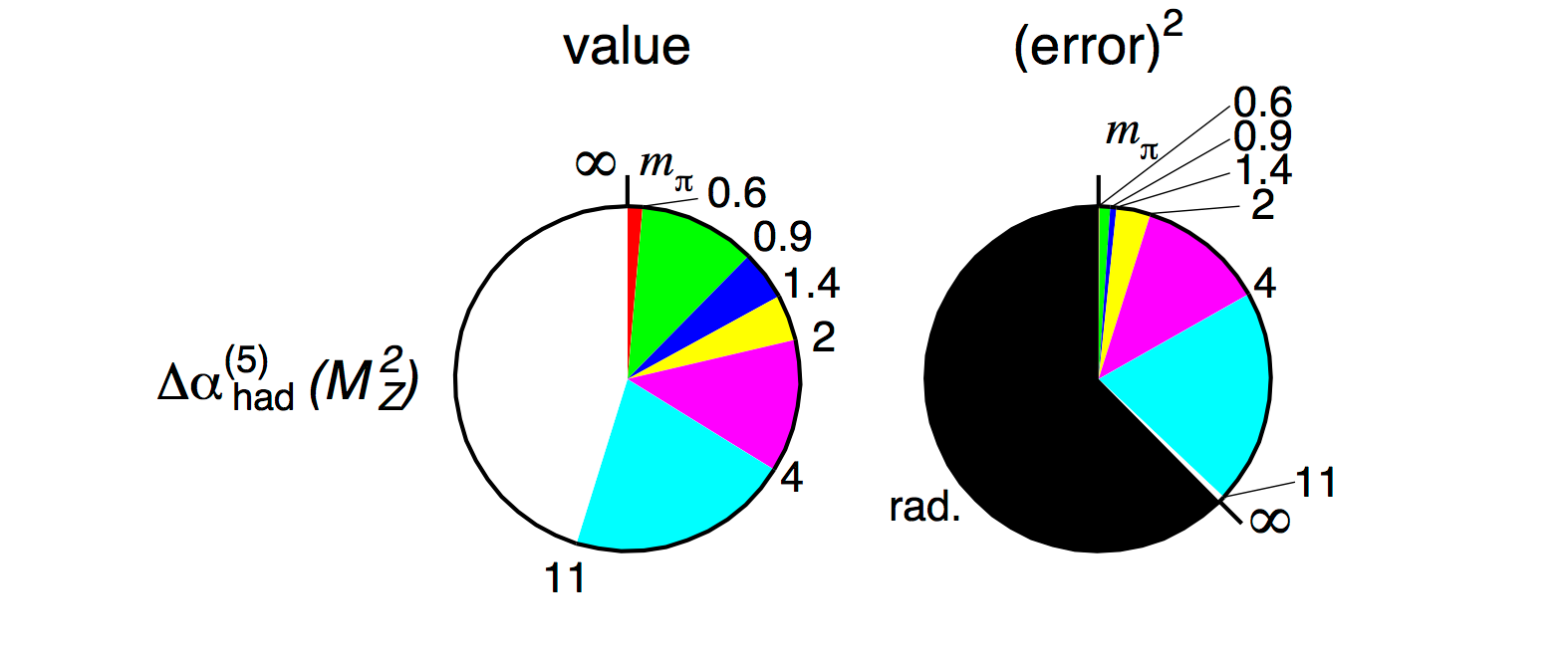}}\hfill
  \caption{Pie charts showing the fractional contributions to the total mean value and (error)$^2$ of both $a_{\mu}^{\rm had, \, LO \, VP}$ and $\Delta\alpha_{\rm had}^{(5)}(M_Z^2)$ from various energy intervals.}  \label{piechart}
\end{figure} 
 \begin{figure}[!t]
\centering
  \subfloat[The hadronic $R$ ratio]{%
    \includegraphics[width= 0.49\textwidth]{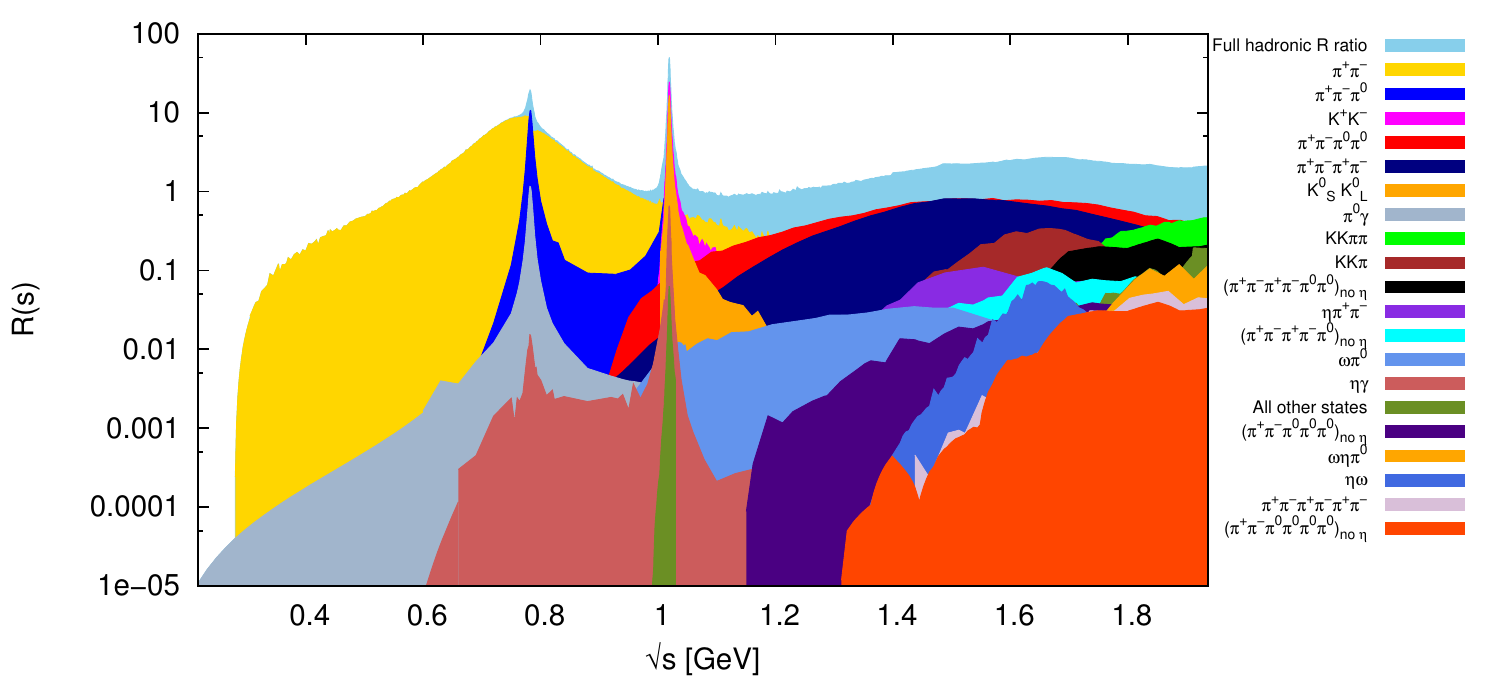}}\hfill
    \vspace{0.5cm}
  \subfloat[The uncertainty of the hadronic $R$ ratio]{%
    \includegraphics[width= 0.49\textwidth]{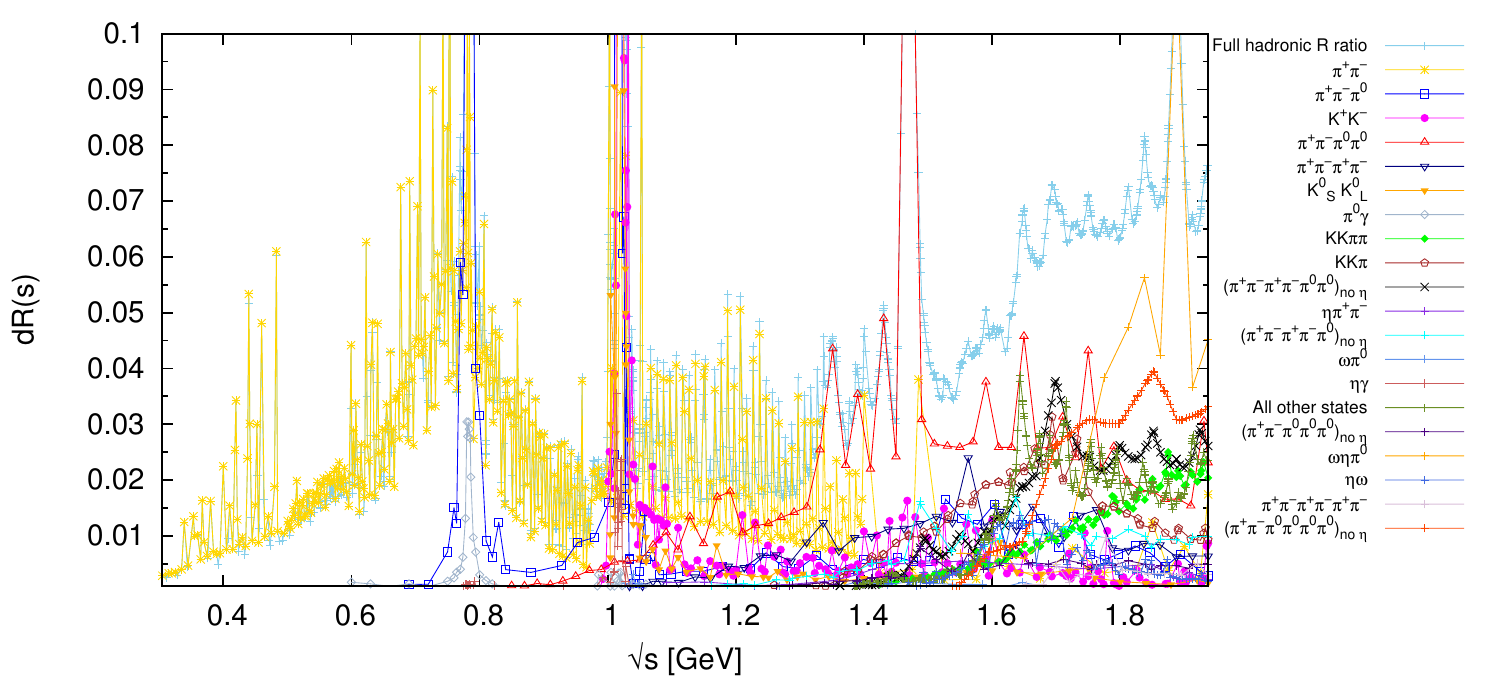}}\hfill
\vspace{-0.5cm}
  \caption{Contributions to the total hadronic $R$ ratio from the different final states (left panel) and their uncertainties (right panel) below 1.937 GeV. The full $R$ ratio and its uncertainty is shown in light blue in each plot, respectively. Each final state is included as a new layer on top in decreasing order of the size of its contribution to $a_{\mu}^{\rm had, \, LO \, VP}$.}  \label{hadxSec}
\end{figure} 
From the sum of all hadronic contributions, the estimate for $a_{\mu}^{\rm had, \, LO \, VP}$ from this analysis is~\cite{Keshavarzi:2018mgv}
\beq\label{LOHVP_KNT18}
a_{\mu}^{\rm had, \, LO \, VP} = (693.27  \pm 1.19_{\rm stat} \pm 2.01_{\rm sys} \pm 0.22_{\rm vp} \pm 0.71_{\rm fsr}) \times 10^{-10} =  (693.27 \pm 2.46_{\rm tot}) \times 10^{-10} \ , 
\eeq
where the uncertainties include all available correlations and local $\chi^2_{\rm min}/{\rm d.o.f.}$ inflation.
Using the same data compilation as for the calculation of $a_{\mu}^{\rm had, \, LO \, VP}$, the next-to-leading order (NLO) contribution to $a_{\mu}^{\rm had, VP}$ is determined here to be $a_{\mu}^{\rm had, NLOVP} =  (-9.82 \pm 0.04) \times 10^{-10}$ . The corresponding result for $\Delta\alpha_{\rm had}^{(5)}(M_Z^2)$ is~\cite{Keshavarzi:2018mgv}
\beq \label{delalphahad_KNT18}
\Delta\alpha_{\rm had}^{(5)}(M_Z^2) = (276.11  \pm  0.26_{\rm stat} \pm 0.68_{\rm sys} \pm 0.14_{\rm vp} \pm 0.82_{\rm fsr}) \times 10^{-4}  = (276.11 \pm 1.11_{\rm tot}) \times 10^{-4} \ ,
\eeq
where the superscript (5) indicates the contributions from all quark flavours except the top quark. The fractional contributions to the total mean value and uncertainty of both $a_{\mu}^{\rm had, \, LO \, VP}$ and $\Delta\alpha_{\rm had}^{(5)}(M_Z^2)$ from various energy intervals is shown in Figure~\ref{piechart}. Figure~\ref{hadxSec} shows the contributions from all hadronic final states to the hadronic $R$ ratio and its uncertainty below 1.937 GeV. 

\subsection{SM prediction of $g-2$ of the muon and $\alpha(M_Z^2)$} \label{g-2muon}

\begin{figure}[!t] 
  \centering
    \includegraphics[width=0.6\textwidth]{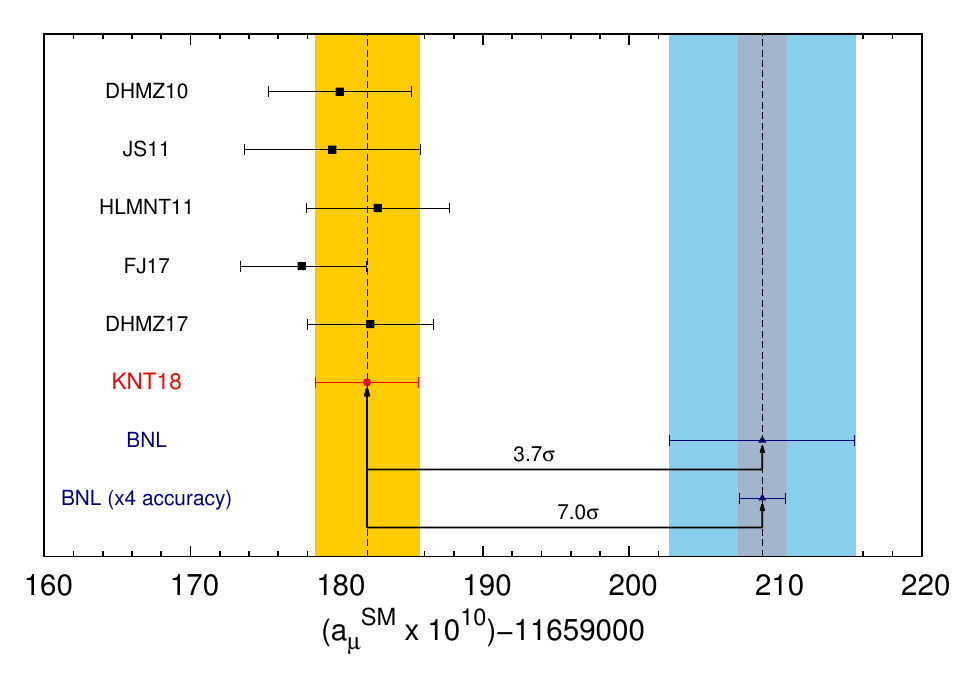}
     \caption{A comparison of recent and previous evaluations of $a_{\mu}^{\rm SM}$. The analyses listed in chronological order are: DHMZ10~\cite{Davier:2010nc}, JS11~\cite{Jegerlehner:2011ti}, HLMNT11~\cite{Hagiwara:2011af}, FJ17~\cite{Jegerlehner:2017lbd} and DHMZ17~\cite{Davier:2017zfy}. The prediction from this work is listed as KNT18~\cite{Keshavarzi:2018mgv}, which defines the uncertainty band that other analyses are compared to. The current uncertainty on the experimental measurement~\cite{Bennett:2002jb,PDG2016} is given by the light blue band. The light grey band represents the hypothetical situation of the new experimental measurement at Fermilab yielding the same mean value for $a_{\mu}^{\rm exp}$ as the BNL measurement, but achieving the projected four-fold improvement in its uncertainty~\cite{Grange:2015fou}.}     \label{amuCompare}
\end{figure}

Combining the results for $a_{\mu}^{\rm had, \, LO \, VP}$ and $a_{\mu}^{\rm had, \, NLO \, VP}$ with the contributions from QED: $a_{\mu}^{\rm QED} = (11 \ 658 \ 471.8971 \pm 0.007)  \times 10^{-10}$~\cite{Aoyama:2012wk}, the electro-weak sector: $a_{\mu}^{\rm EW}  = (15.36 \pm 0.10) \times 10^{-10}$~\cite{Gnendiger:2013pva}, the hadronic vacuum polarisation at NNLO: $a_{\mu}^{\rm had, \, NNLO \, VP}  =  (1.24 \pm 0.01) \times 10^{-10}$~\cite{Kurz:2014wya}, the hadronic light-by-light (LbL) at LO: $a_{\mu}^{\rm had, \, LbL} =  (9.8  \pm 2.6) \times 10^{-10}$ ~\cite{Nyffeler:2016gnb} and the hadronic LbL at NLO:  $a_{\mu}^{\rm had, \, NLO \, LbL}  =  (0.3  \pm 0.2) \times 10^{-10}$~\cite{Colangelo:2014qya}, the SM prediction of the anomalous magnetic moment of the muon is found to be
\beq \label{amuSMfinal}
a_{\mu}^{\rm SM}  =  (11\ 659 \ 182.05  \pm 3.56) \times 10^{-10} \, .
\eeq
Comparing this with the current experimental measurement results in a deviation of $\Delta a_{\mu} = (27.05 \pm 7.26)\times 10^{-10}$, corresponding to a $3.7\sigma$ discrepancy. This result is compared with other determinations of $a_{\mu}^{\rm SM}$ in Figure~\ref{amuCompare}. The total value of the QED coupling at the $Z$ boson mass is found in this work to be
\beq
\alpha^{-1}(M_Z^2) = \Big(1-\Delta\alpha_{\rm lep}(M_Z^2)-\Delta\alpha_{\rm had}^{(5)}(M_Z^2)-\Delta\alpha_{\rm top}(M_Z^2)\Big)\alpha^{-1} =  128.946 \pm 0.015\, .
\eeq

\section{Conclusions}\label{Conclusions}

This analysis, KNT18~\cite{Keshavarzi:2018mgv}, has completed a full re-evaluation of the hadronic vacuum polarisation contributions to the anomalous magnetic moment of the muon, $a_{\mu}^{\rm had, \, VP}$ and the hadronic contribution to the effective QED coupling at $Z$ boson mass, $\Delta\alpha_{\rm had}(M_Z^2)$. Combining all available $e^+e^- \rightarrow {\rm hadrons}$ cross section data, this analysis found $a_{\mu}^{\rm had, \, LO \, VP} = (693.27 \pm 2.46)\times 10^{-10}$ and $a_{\mu}^{\rm had, \, NLO \, VP} = (-9.82 \pm 0.04)\times 10^{-10}$. This has resulted in a new estimate for the Standard Model prediction of $a_{\mu}^{\rm SM}  =  (11\ 659 \ 182.05  \pm 3.56) \times 10^{-10}$, which deviates from the current experimental measurement by $3.7\sigma$. 

\section*{Acknowledgements}

We would like to thank the organisers of {\em LFC17: Old and New Strong Interactions from LHC to Future Colliders} for a very productive and enjoyable workshop.

\let\oldthebibliography\thebibliography
\let\endoldthebibliography\endthebibliography
\renewenvironment{thebibliography}[1]{
  \begin{oldthebibliography}{#1}
    \setlength{\itemsep}{0.4em}
    \setlength{\parskip}{0em}
}
{
  \end{oldthebibliography}
}

\end{document}